# Low power communication signal enhancement method of Internet of things based on nonlocal mean denoising

Mingchuan TIAN,[1] JizhengLiu,[2]

1 Nanyang Technological University, 50 Nanyang Ave, 639798, Singapore
2 Beijing Institute of Technology, Beijing, 100081, China

Correspondence should be addressed to JizhengLiu; liujz_ev@bit.edu.cn

**Abstract:** In order to improve the transmission effect of low-power communication signal of Internet of things and compress the enhancement time of low-power communication signal, this paper designs a low-power communication signal enhancement method of Internet of things based on nonlocal mean denoising. Firstly, the residual of one-dimensional communication layer is pre processed by convolution core to obtain the residual of one-dimensional communication layer; Then, according to the two classification recognition method, the noise reduction signal feature recognition of the low-power communication signal of the Internet of things is realized, the non local mean noise reduction algorithm is used to remove the low-power communication signal of the Internet of things, and the weight value between similar blocks is calculated according to the European distance method. Finally, the low-power communication signal enhancement of the Internet of things is realized by the non local mean value denoising method. The experimental results show that the communication signal enhancement time overhead of this method is low, which is always less than 2.6s. The lowest bit error rate after signal enhancement is about 1%, and the signal-to-noise ratio is up to 18 dB, which shows that this method can achieve signal enhancement.

## 1 Introduction

In recent years, cognitive radio signal communication technology has shown great vitality and market potential in a large range of the world. However, in the process of signal transmission, due to the multipath effect in the channel, the limitation of channel transmission bandwidth and the imperfection of channel transmission characteristics, the signal will be seriously affected by all kinds of noise and electromagnetic interference, which will directly degrade the quality of communication and affect the reception and analysis of the signal at the receiver [1].For example, reduce spectrum sensing, increase demodulation difficulty, etc. Only by obtaining high-quality signals, further analysis and processing are meaningful. Therefore, removing noise and interference in communication signals, that is, signal enhancement, is an important technology to promote the development of wireless communication field. Therefore, researchers have proposed many signal enhancement methods, which can be divided into linear method and nonlinear method [2]. The linear signal enhancement method is relatively simple, but it still has low performance for nonlinear signals, and can not find the global optimal solution for noise elimination. In addition, because these methods are based on the assumption that the signal is stationary, their effectiveness is generally acceptable, and the actual signal usually has non-stationary statistical characteristics [3-5].Nonlinear methods have become a research hotspot in recent two decades because they can clarify the spectrum and time information in the signal at the same time. However, the traditional signal enhancement methods use the prior information of noise or interference to map to the separable transform domain for separation, such as band-pass filtering. However, due to the random characteristics of noise and interference, the artificial construction of the corresponding separable transform domain often has a strong a priori to noise and interference. Therefore, at present, there is no more complete method to adaptively enhance the signal of time-varying system. For this reason. Relevant scholars have conducted comparative research and made some progress.

Zhong Nan et al. Proposed a satellite navigation signal enhancement method in the tunnel based on virtual satellite [6]. By establishing the signal propagation model, simulate the navigation signal of low elevation satellite located in the extension direction of the tunnel at both ends of the tunnel, and receive the solution in the tunnel after sending it to realize positioning. At the same time, the signal delay control method is used to pre compensate the pseudo range error. Through the experimental analysis of the hardware system, its positioning ability in the tunnel is verified. The simulation results show that the positioning accuracy of the system meets the needs of most tunnels. Cao Pengyu et al. Proposed an LPI radar signal enhancement method based on dae-gan network [7], combined with the advantages of noise reduction self encoder and generation of countermeasure network, constructed a noise enhancement network and signal enhancement network for countermeasure training. The noise enhancement network doped more complex noise components into the noisy signal, and the signal enhancement network reduced the noise components in the noisy signal as much as possible. In this paper, dae-gan network is used to realize complex high-dimensional noise reduction. This method can effectively improve the effect of signal enhancement, but the signal enhancement takes a long time.

To solve the above problems, this paper designs a low-power communication signal enhancement method of the Internet of things based on non local mean denoising. The low-power communication signal of the Internet of things is preprocessed through the deep residual network in deep learning, and the low-power communication signal enhancement of the Internet of things is realized according to the non local mean denoising method.

## 2 Low power communication signal preprocessing of Internet of things based on deep learning

## 2.1 Depth residual network analysis

Deep neural network always has the problem of network degradation. Generally speaking, with the increase of the depth of the network, the classification performance of the network should be stronger and stronger. However, the current situation is that the performance of the network gradually tends to saturation with the rise of the network depth, but will decline rapidly after the depth rises to a certain node. This phenomenon is called network degradation [8]. For the network whose accuracy is close to saturation, if the identity mapping layer is added in the network structure to make its input equal to the output, it will not increase the error after propagation while continuing to increase the network depth, that is, deepening the network through this method will not increase the training error. The design method of residual network comes from this [9-12]. The residual network is proposed mainly to reduce the network degradation caused by the rise of network depth. The solution is to construct a "residual element". The structure of residual element is shown in Figure 1.

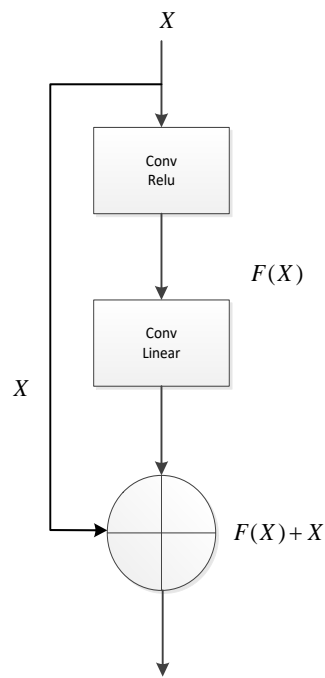

Figure 1: residual element

For a neural network, suppose that the input of a certain section in the network is $X$ and its corresponding expected output is $Z(X)$, that is, $Z(X)$ is the expected potential mapping. Generally, when the network is deepened, the training difficulty will increase [13]. In the residual network structure diagram in the figure above, a path from input to output is added on the basis of network mapping, the input $X$ is directly transferred to the output as the initial result, and the output result is $Z(X) = F(X) + X$. When $F(X) = O$, there is $Z(X) = X$, that is, the identity mapping that will not increase the error mentioned earlier [14]. RESNET changes the learning strategy. For an input $X$, it no longer learns its expected mapping $Z(X)$ directly through the convolution layer [15], but learns the expected residual mapping through the network, that is, $Z(X)-X$. Compared with the expected mapping, the residual mapping is easier to optimize and deepens the network depth on the premise of avoiding network degradation.

**2.2 Impulse noise preprocessing**

The INP used in this paper can be regarded as the neutralization of truncation and zeroing in the threshold suppression method. The difference lies only in the processing of the part whose amplitude is higher than the threshold. Simulation experiments show that it has better pulse suppression effect than the two methods. The main task of INP is to suppress the nonlinear part of the received signal $y(t)$ whose amplitude is greater than the threshold $\tau_r$. The output signal can be expressed as:

$$y_{non}(n) = \begin{cases} y(n), & |y(n)| \leq \tau_r \\ y(n)(\dfrac{\tau_r}{|y(n)|})^2 & |y(n)| > \tau_r \end{cases} \quad (1)$$

Wherein, $\tau_r$ can be obtained from formula (2)

$$\tau_r = (1+2\tau_0)\tau_Q \quad (2)$$

Where, $\tau_0$ is the constant coefficient, set to 1.5, and $\tau_Q$ is the second quartile value of the received signal $y(n)$ modulus $|y(n)|$ [16]. After pulse suppression, the signal needs to be further normalized to obtain the final output signal of INP, that is, the input of RCGAN noise reduction network:

$$y_p(n) = \frac{y_{non}(n)}{\max(|y_{non}(n)|)} \quad (3)$$

In order to improve the ability of the network to retain the signal details in the process of noise reduction, the middle layer of the generator adopts the extended convolution structure widely used in the field of image semantic segmentation [17]. This structure expands the receptive field of the convolution kernel by inserting zeros in the convolution kernel. Compared with the receptive field expansion methods such as down sampling used in the standard convolution structure, this method has stronger retention ability of detailed information. Generally, the expansion rate is used to represent the interval of adjacent elements in the convolution kernel, and the expansion rate of the standard convolution kernel is 1. The expansion rate $r$ of three one-dimensional expansion convolution layers in RCGAN generator is taken as 1, 2 and 4 respectively, and the length of convolution core is 3. The perceived field of view of convolution cores in different layers can be expressed as:

$$F_r = 2^{r+1} - 1 \quad (4)$$

Based on this, it can be calculated that the convolution kernel receptive field sizes of the three one-dimensional expanded convolution layers are 3, 7 and 15 respectively.

Figure 2 compares the change process of receptive visual field of three one-dimensional standard convolution layers and three one-dimensional expanded convolution layers when the convolution kernel length is 3.

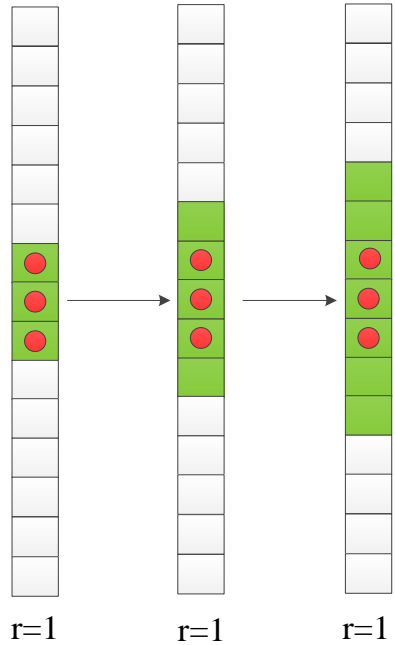

r=1　　　r=1　　　r=1

(a) One dimensional standard convolution

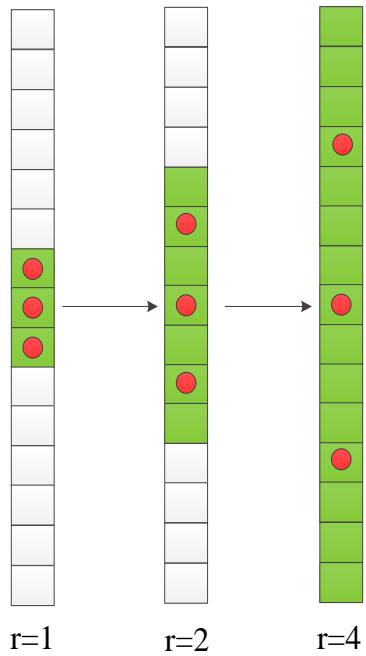

r=1　　　r=2　　　r=4

(b) One dimensional extended convolution

Figure 2:　Comparison of receptive field between one-dimensional standard convolution layer and expanded convolution layer

The solid circle in Figure 2 represents the position of non-zero value in the convolution core, and the solid square represents the receptive field of view of the convolution core. The receptive field of vision showed a linear growth trend, while the receptive field of vision showed a linear growth trend. The research shows that expanded convolution achieves better retention of small-scale information features in semantic segmentation by virtue of this information lossless sensory field expansion method. Therefore, we use this structure in the middle layer of the generator to improve the network's ability to extract signal details.

**2.3 Signal generation method**

The flow chart of signal noise reduction algorithm generated by low-power communication of Internet of things used in this paper is shown in Figure 3:

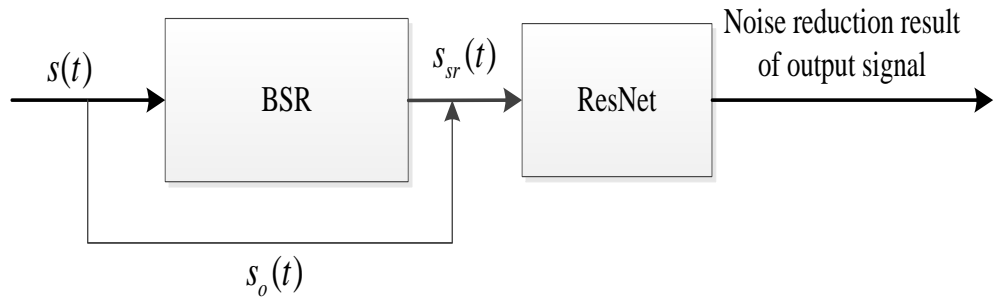

Figure 3: signal denoising algorithm processing flow

The first part is the bistable stochastic resonance (BSR) system. After the input signal $s_{sr}(t)$ passes through the system, the output signal $s_{sr}(t)$ is obtained. Then, the data set is written together with the non resonant signal $s_o(t)$ and input into the depth residual network (RESNET), and finally the noise reduction result is output. After the modulated signal is generated, the bistable coefficient is determined, and then the signal through stochastic resonance is output through the BSR system. After each signal sample is generated, in order to make the signal waveform correspond one by one, after each original sample is generated, the sample is generated through the BSR system, and then the samples before and after resonance are stored in the data set at the same time. And mark whether the signal resonates. The construction process is shown in Figure 4.

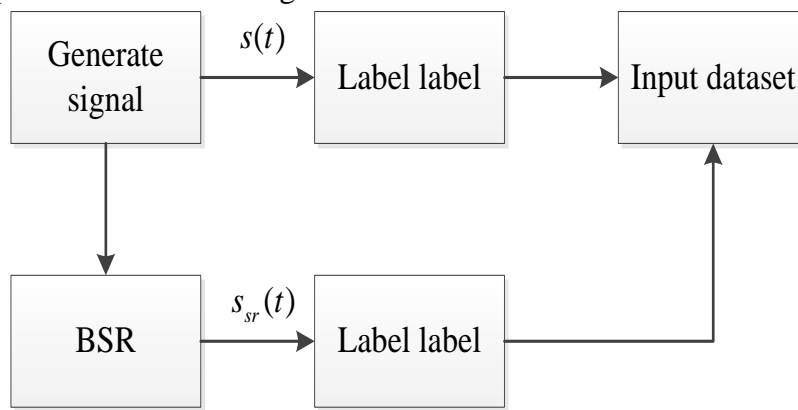

Figure 4: data set construction method

After the signal is generated, label the two types of signals, and then input RESNET for noise reduction. After the signal noise reduction is completed, this paper continues to build a binary classification network C based on CNN to automatically extract the detection features used to characterize the presence or absence of the signal in the noise reduction signal and complete the binary classification recognition. With the characteristics of weight sharing and local perception field, CNN has obvious advantages in local feature extraction while greatly reducing the amount of model parameters. It has been widely used in many data classification problems. After INP preprocessing and effective noise reduction of Rcgan network, the noise components in the received signal are significantly suppressed, while the useful signal components will be well preserved. Then input the signal waveform after noise reduction into the CNN classifier. Due to the significant improvement of signal-to-noise ratio, the difficulty of signal detection after noise reduction will be greatly reduced, and the false alarm rate will be significantly reduced.

In the structure setting of classifier C, this paper similarly adopts the step convolution structure to compress the feature dimension. Its overall structure is similar to D. the convolution layer has six layers, including 16, 32, 64, 128, 256 and one convolution core respectively. At the end of the convolution layer, it is connected with the softmax classifier through a full connection layer to output the probability of communication signal and pure noise in the received signal respectively, The setting of the activation function is the same as that of the decider. When $(y_s, y_p, y_L)$ is used to represent the transmission data in a training sample, the INP preprocessed data and the real tag value after one hot coding, the predicted tag probability vector output by the softmax classifier $\hat{y}_L$ can be expressed as:

$$\hat{y}_L = \begin{bmatrix} P(L_p = 0 | Cout) \\ P(L_p = 1 | Cout) \end{bmatrix} = \frac{1}{\sum_{i=0}^{1} \exp(Cout_i)} \begin{bmatrix} \exp(Cout_0) \\ \exp(Cout_1) \end{bmatrix} \quad (5)$$

$$Cout = f_{CNN}(G(y_p)) \quad (6)$$

Where, $L_p$ represents the predicted tag value, $Cout$ represents the input vector of softmax, and $f_{CNN}$ represents the nonlinear function formed by the network before softmax layer in classifier C.

## 3 Non local mean denoising and enhancement method of low power communication signal in Internet of things

Using a large number of redundant information with similar structure in the natural image, make full use of the redundant information on these images to reduce the noise of the image[18]. When processing each pixel in the noisy image, the distribution around the pixel will be evaluated and compared, and the difference similarity of the distribution will be used to calculate the weight $\omega$. The performance of non local mean denoising algorithm is better than other traditional image denoising algorithms, and the denoising effect is better[19-21].

For a noisy image $Y(i) = X(i) + n(i)$, $X(i)$ is the original image and $n(i)$ is noise. Take the noise reduction of a pixel on a noisy image as an example. First, take this pixel as the center point, and then create a neighborhood window[22]. Next, traverse the whole noisy image to find similar blocks with similar neighborhood window structure. Then, the weighted re equalization of these similar blocks is calculated, and the calculated pixels are the pixels after noise reduction[23-25]. The formula of nonlocal mean noise reduction algorithm is shown in formula (7):

$$NLM[\overline{X}](i) = \sum_{j \in I} w(i,j) Y(j) \quad (7)$$

Suppose that the noisy signal Y can be composed of pure signal s and additive interference d:

$$y = s + d \quad (8)$$

Under the set conditions, y obtains the estimated value $\hat{s}$ of s, which is the main principle of signal enhancement. In the form of short-time Fourier transform (STFT), $Y_{n,k} \exp(j\alpha_n, k)$, $S_{n,k} \exp(j\varphi_n, k)$ and $\hat{S}_{n,k} \exp(j\hat{\varphi}_n, k)$ represent y, s and $\hat{s}$ in the n-th frame respectively, and the frequency sequence number is represented by K = 1, 2,..., K. without considering the phase information, the main purpose of signal enhancement task is to minimize its error function:

$$E_r = \sum_{k=1}^{K} \left( \hat{S}_{n,k} - S_{n,k} \right)^2 \quad (9)$$

Let the amplitude spectrum vector and estimated value of the pure signal on the nth frame be represented by and respectively. At this time, the function error can be expressed as:

$$E_r = \left\| \hat{S}_n - S_n \right\|_2^2 \quad (10)$$

The research on the signal enhancement method of deep neural network hospital wireless communication network can be understood as: using the training parameter set $\theta$ to build a nonlinear function $f_\theta$, which requires that the function $f_\theta$ must have cumbersome characteristics, so it is used to ensure the new error function:

The research on the signal enhancement method of wireless communication network can be understood as: using the training parameter set $\theta$ to build a nonlinear function $f_\theta$, which requires that the $f_\theta$ function must have cumbersome

characteristics, so it is used to ensure the new error function:

$$E_r = \|f_\theta(X_n) - S_n\|_2^2 \quad (11)$$

Minimum to obtain the target output:

$$\hat{S}_n = f_\theta(X_n) \quad (12)$$

Where, $X_n = [Y_{n-N}, Y_{n-N+1}, \cdots, Y_n, \cdots, Y_{n+N-1}, Y_{n+N}]$ is the training feature of the nth frame, which is composed of half [(2n + 1) frame] amplitude spectrum vector of the nth frame, (2n + 1) is the input length.

When looking for the pixel block of the neighborhood window, we usually determine the weight value $w(i, j)$ by calculating the Euclidean distance between similar blocks, and the Euclidean distance $d(i, j)$ between similar blocks, as shown in formula (13):

$$d(i, j) = \|Y(N_i) - Y(N_j)\|_{2,a}^2 \quad (13)$$

The weight value $w(i, j)$ is calculated as shown in formula (14):

$$w(i, j) = \frac{1}{Y(i)} e^{-d(i,j)^2/h^2} \quad (14)$$

$h$ is the smoothing control parameter of nonlocal mean noise reduction, which has a very important impact on the final noise reduction effect. $\overline{X}$ is the image after nonlocal mean noise reduction. The one-dimensional off diagonal slice $c_{4x}(m)$ of the fourth-order cumulant of IOT low-power communication signal $x(n)$ can simultaneously resist Gaussian noise and maintain the basic framework of available signal $s(n)$. Therefore, based on the weak signal $x(n)$, the estimated data $\hat{c}_{4x}(m)$ based on the one-dimensional off diagonal slice of the fourth-order cumulant $x(n)$ can be calculated, and then $\hat{c}_{4x}(m)$ pairs of fir (finite impulse response) filters can be used to create and collect the signal from the noise. The impulse response of FIR filter is defined as the following formula:

$$h(m) = \begin{cases} \hat{c}_{4x}(L-m), m = 0, 1, \ldots, L \\ \hat{c}_{4x}(m-L), m = L+1, L+2, \ldots, 2L \end{cases} \quad (15)$$

Among them, the maximum lag number is expressed in $L$, and the length of impulse response of FIR filter is expressed in $2L+1$. Because $\hat{c}_{4x}(m)$ is similar to $s(n)$ to some extent, it can be concluded that the FIR filter defined in the above formula is the relative matched filter of the available signal $s(n)$, and its output signal-to-noise ratio tends to be idealized. The output formula of the filter is as follows:

$$y(n) = \gamma \sum_{m=0}^{2L} h(m) x(n-m) \qquad (16)$$

In the above formula, the gain factor is expressed as $\gamma$, which has the control function of filter output, and the value is usually the reciprocal of kurtosis coefficient [26-27]. At this time, the output result of the filter is the enhanced low-power communication signal of the Internet of things, which completes the enhancement of the low-power communication signal of the Internet of things.

## 4 Experiment

### 4.1 Experimental Design

In reality, most of the environmental background noise is unstable signals. In order to ensure the enhancement effect of low-power communication signals of the Internet of things, the noise needs to be adjusted in time. In this paper, MSP430 series single chip microcomputer with hardware multiplier is selected, so that it can not be disturbed by CPU in the calculation process; The A/D converter adopts a 12 bit, 60K ROM; The noise environment background of the simulation experiment is $\geq -5dB$.

### 4.2 Analysis Of Experimental Results

#### 4.2.1 Time domain waveform analysis of enhancement effect

Through the time domain waveform, the simulation results of this method and reference [6] method are analyzed and compared. The low-power communication signal acquisition frequency of the experimental Internet of things is set to 8kHz. In the low-power communication signal enhancement method of the Internet of things in this paper, the filter order of the first two stages is 32, and the input signal of the first stage is divided into three output channels by the filter according to the frequency factor. The experimental results are shown in the following time domain waveform diagram, as shown in Figure 5:

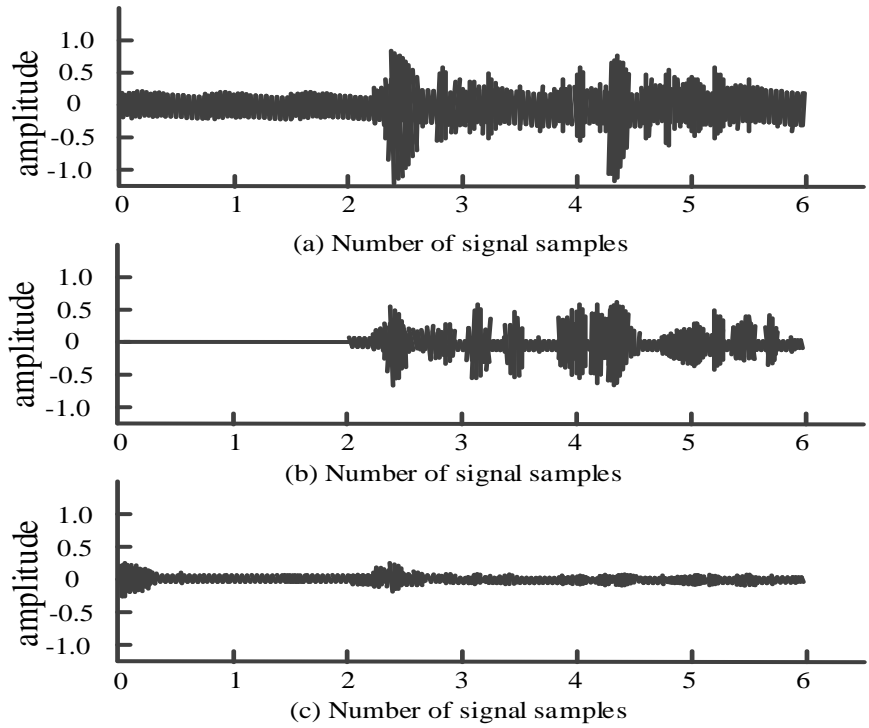

Figure 5: time domain waveform

The y-axis is the amplitude of the dimensionless processed signal, and the x-axis is the number of samples of the signal, about 60000. (a) in the above figure is a signal with noise, and the interference intensity of the noise is very high. After calculation, the signal-to-noise ratio of the signal to the noise is 1.121 dB. Therefore, the weak signal of the original communication is very unclear and is submerged by the noise in a large range. Figure (b) shows the time-domain waveform of the method of reference [6]. Although the algorithm can cancel the output signal, the output signal-to-noise ratio is greatly reduced to 0.544 dB. It can be seen that this method is immune to noise and low-power communication signals of the Internet of things, resulting in signal distortion. It can not reduce noise or enhance the signal alone. Figure (c) shows the second stage output of the method in this paper. The signal-to-noise ratio of the signal is increased to about 10 times of the original signal, and its value is about 13.465 dB, which greatly enhances the low-power communication signal of the Internet of things. At the same time, the noise is well suppressed and removed.

**4.2.2 Signal enhancement time**

The experiment further analyzes the time cost of low-power communication signal enhancement of the Internet of things by the method in this paper, the method in reference [6] and the method in reference [7]. The results are shown in Table 1:

Table 1: time cost of signal enhancement by different methods (s)

| Number of signals / piece | Paper method | Reference [6] method | Reference [7] method |
|---|---|---|---|
| 1000 | 0.2 | 9.2 | 6.9 |
| 1500 | 0.8 | 16.2 | 15.2 |
| 2000 | 0.9 | 19.9 | 19.5 |
| 2500 | 1.2 | 22.6 | 23.1 |
| 3000 | 1.8 | 26.9 | 28.6 |
| 3500 | 2.0 | 32.1 | 33.0 |
| 4000 | 2.2 | 38.9 | 38.6 |
| 4500 | 2.3 | 42.9 | 40.8 |
| 5000 | 2.6 | 46.8 | 45.9 |

By analyzing the experimental data in Table 1, it can be seen that the time cost of IOT low-power communication signal enhancement has changed with the number of IOT low-power communication signals enhanced by the methods in this paper, reference [6] and reference [7]. Among them, the time cost of low-power communication signal enhancement of the Internet of things in this method is low and always less than 2.6s. The time cost of low-power communication signal enhancement of the Internet of things in reference [6] method is low in the early stage, but with the increase of the number, the iteration time continues to increase. The time cost of low-power communication signal enhancement of the Internet of things in reference [7] method is always higher than the first two methods. It can be seen that this method has shorter enhancement time and certain work efficiency.

**4.3.2 Signal enhancement effect**

In the experiment, firstly, the bit error rate after weak signal enhancement in wireless communication network under electromagnetic interference environment is analyzed, and the bit error rate after signal enhancement using this method, reference [4] method and reference [5] method is compared. The results are shown in Figure 6.

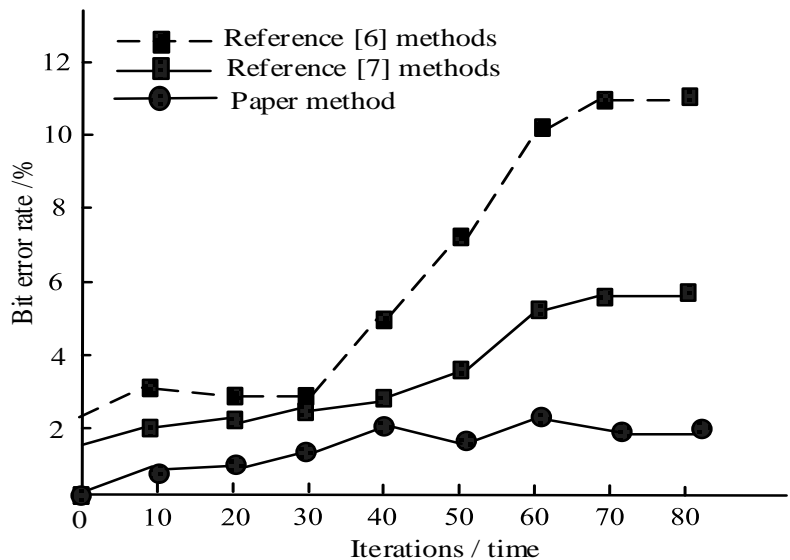

Figure 6: bit error rate analysis of weak signal enhanced by different methods

By analyzing the experimental results in Figure 6, it can be seen that there are some differences in the bit error rate after signal enhancement using the methods in this paper, reference [6] and reference [7]. Among them, the bit error rate of this method after signal enhancement is lower than that of reference [7] and reference [6], and the fluctuation trend of its curve is small, the lowest is about 1%, while the experimental curve of bit error rate of the other two methods after signal enhancement fluctuates greatly and is always higher than that of this method, so it can be seen that the bit error rate of this method is lower. This is because this method takes into account the interference of electromagnetic wave during signal enhancement in communication, calculates the autocorrelation function between random signals, maintains the stability of the signal through the average of sample sequence time, adjusts the frequency offset and initial phase through low-pass filter by amplitude modulation coefficient, and estimates the amplitude of weak signal and the interference degree of signal by maximum likelihood, The signal enhancement is realized.

### 4.3.3 Signal to Noise Ratio

For the reference [6] method with good effectiveness and accuracy, the weak signal enhanced by this method has better effectiveness, lower distortion and stronger noise reduction ability.

After the method in reference [6] is coherently averaged $N$ times, the optimization degree of its signal-to-noise ratio will be increased by $\sqrt{N}$ times. This method can not only idealize the noise reduction of the environmental background, but also improve the optimization progress of signal-to-noise ratio, highlight the available signals and enhance them adaptively. The figure 7 shows the trend change of signal-to-noise ratio in the enhancement stage, of which $N=150$, $\mu=5e-6$.

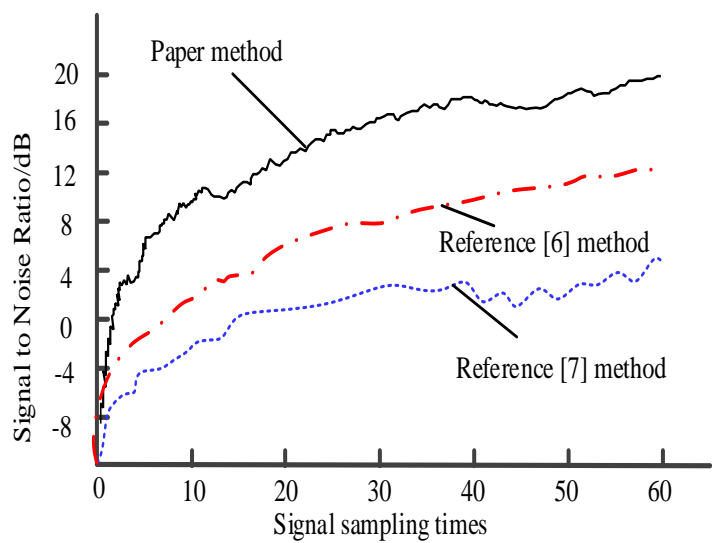

Figure 7 Comparison of signal-to-noise ratio under different methods

According to the analysis of Figure 7, the signal-to-noise ratio is different under different methods. When the signal is sampled 10 times, the signal-to-noise ratio of the method in reference [6] is 2 dB, the signal-to-noise ratio of the method in reference [7] is 1 dB, and the signal-to-noise ratio of the method in this paper is 11 dB; When the signal is sampled 50 times, the signal-to-noise ratio of reference [6] method is 10 dB, the signal-to-noise ratio of reference [7] method is 2 dB, and the signal-to-noise ratio of this method is 18 dB; It can be seen from the above figure that the signal-to-noise ratio of this method continues to grow, and its curve trend has been higher than the other two traditional methods, indicating that this method has very significant advantages.

**4.3.4 signal enhancement stability test**

After the signal is enhanced, the signal gain coefficient can reflect the enhancement effect of the signal enhancement method. Set the gain coefficient of the signal as $\alpha$ and the optimal gain interval as $[0,1]$. The higher the test result of the signal gain coefficient in this interval, the better the enhancement effect of the signal, and vice versa. The methods proposed in this paper, reference [5] and reference [6] are used for signal enhancement, and the signal gain coefficients of the three methods are tested. The test results are shown in Figure 8.

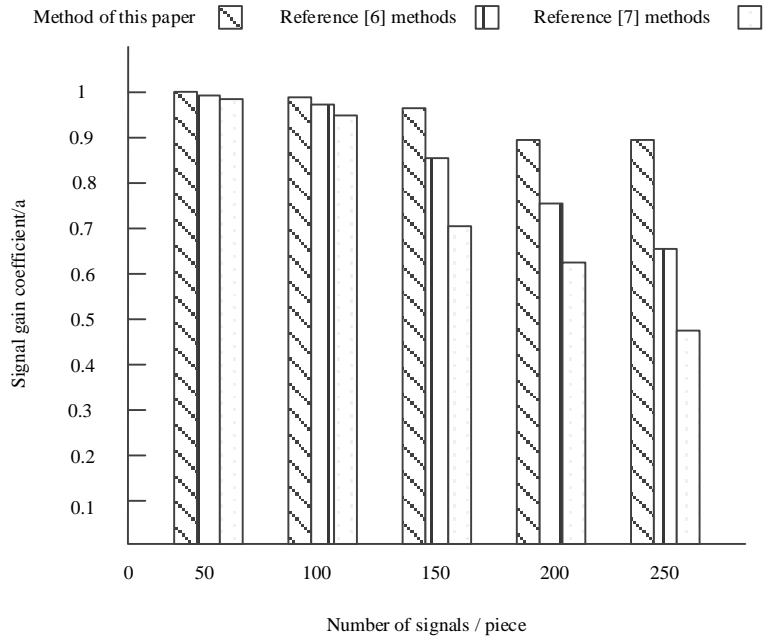

Figure 8: test results of signal gain coefficient

As can be seen from Figure 8, the increase in the number of signals will reduce the gain coefficient after signal enhancement. It can be seen from the analysis of Figure 5 that the method proposed in this paper will reduce the test results with the increase of signals, but when the number of signals increases to a certain range, the method proposed in this paper can stabilize the test results within a certain coefficient. The test results of reference [6] method and reference [7] method are similar at the initial stage of the test, but with the progress of the test, the gap between them continues to widen. Finally, the test results of reference [6] method are much higher than those of reference [7]. It can be seen that the gain coefficient measured by the method proposed in this paper is high after signal enhancement, which shows that the enhancement effect of this method is good.

## 5 Conclusion

This paper designs a low-power communication signal enhancement method for the Internet of things based on nonlocal mean denoising. Preprocess the low-power communication signal noise of the Internet of things through the deep residual network, realize the noise reduction signal feature recognition of the low-power communication signal of the Internet of things according to the two classification recognition method, calculate the weight value between similar blocks according to the Euclidean distance method, and enhance the low-power communication signal of the Internet of things through the non local mean denoising method. The experimental results show that the communication signal enhancement time overhead of this method is low, always less than 2.6s, the lowest bit error rate after signal enhancement is about 1%, and the highest signal-to-noise ratio is 18dB, which shows that this method has very significant advantages in signal enhancement.

# References

[1] Tang Minglei, Zhang Wenpeng, Jiang Weidong, Gao Xunzhang. Micro motion signal enhancement method based on multi-resolution saliency filtering [J]. Systems engineering and electronic technology, 2022,44 (04): 1148-1157.
[2] Y Ma, Gao M , Wang L , et al. Accuracy Enhancement of Moments-based OSNR Monitoring in QAM Coherent Optical Communication[J]. IEEE Communications Letters, 2020, 24(99):821-824.
[3] D Shukla, Prakash A, Tripathi R. Adaptive Modulation and Coding for Performance Enhancement of Vehicular Communication[J]. Wireless Personal Communications, 2021，26(7)：126-131.
[4] Kim H B , Morris J , Miyashiro K , et al. Astrocytes promote ethanol induced enhancement of intracellular Ca 2+ signals through intercellular communication with neurons[J]. iScience, 2021，18(23):1024306-1024310.
[5] MM Hassan, Ahmed K, Paul B K , et al. Anomalous birefringence and nonlinearity enhancement of As2S3 and As2S5 filled D-shape fiber for optical communication[J]. Physica Scripta, 2021, 96(11):115501-115527.
[6] Zhong Nan, Song Maozhong, Liu Haokai. Satellite navigation signal enhancement method in tunnel based on virtual satellite [J]. Telecommunication technology, 2020,60 (05): 511-516.
[7] Cao Pengyu, Yang Chengzhi, Shi Limeng, Wu Hongchao. LPI radar signal enhancement based on dae-gan network [J]. Systems engineering and electronic technology, 2021,43 (09): 2493-2500.
[8] Zou D, Ma S. Satellite Navigation and Communication Integration Based on Correlation Domain Indefinite Pulse Position Modulation Signal[J]. Wireless Communications and Mobile Computing, 2021, 25(09):1-11.
[9] Moualeu J M, Ngatched T. Physical-Layer Security Enhancement via Relay-Aided D2D Communications Underlaying Cellular Networks[J]. IEEE Open Journal of the Communications Society, 2020, 15(09):413-427.
[10] Younus O I, Hassan N B , Ghassemlooy Z, et al. Data Rate Enhancement in Optical Camera Communications Using an Artificial Neural Network Equaliser[J]. IEEE Access, 2020, 8(36):42656-42665.
[11] Du X, Wu D. Visual inspection system for trackside communication and signal infrastructure[J]. Proceedings of the Institution of Mechanical Engineers Part F Journal of Rail and Rapid Transit, 2020, 235(1):409-416.
[12] Ramadan K , Dessouky M I , El-Samie F . Equalization and Co-Carrier Frequency Offsets Compensations for UWA-OFDM Communication Systems[J]. Wireless Personal Communications,

2022,21(08):1-17.

[13] Ramadan K, Dessouky M I , El-Samie F . A Modified OFDM Configuration with Equalization and CFO Compensation for Performance Enhancement of OFDM Communication Systems[J]. AEU - International Journal of Electronics and Communications, 2020, 126(12):172-179.

[14] Ramadan K, Dessouky M I , El-Samie F . A Modified OFDM Configuration with Equalization and CFO Compensation for Performance Enhancement of OFDM Communication Systems[J].AEU-International Journal of Electronics and Communications, 2020, 126(12):161-173.

[15] Eid M, Sorathiya V, Lavadiya S , et al. Free space and wired optics communication systems performance improvement for short- range applications with the signal power optimization[J]. Journal of Optical Communications, 2021,16(15):139-146.

[16] ZHANG Qichao, MENG Xin, MA Shexiang. Combined Estimation of AIS Mixed Signal Separation and Detection Based on PSP[J]. Computer Simulation, 2021, 38(1):464-470.

[17] Towliat M, Rajabzadeh M, Tabatabaee S . On the Noise Enhancement of GFDM[J]. IEEE Wireless Communication Letters, 2020, 12(19):1-10.

[18] Kamimura R . SOM-based information maximization to improve and interpret multi-layered neural networks: From information reduction to information augmentation approach to create new information[J]. Expert Systems with Application, 2019, 125(5.):397-411.

[19] Bo, L. , Lv, J. , Luo, X. , Wang, H. , & Wang, S. . A novel and fast nonlocal means denoising algorithm using a structure tensor. Journal of supercomputing,2019, 75(2): 770-782.

[20] Amutha, B. , et al. "Enhanced development of communication between the network and the end user by eliminating the interference signals in MIMO channel." Transactions on Emerging Telecommunications Technologies .2020,70(12):: 77-83.

[21] Liu C , Jin H E , Zhang C , et al. Pattern Recognition Method for Time-domain Waveform Images of GIS Partial Discharge[J]. Proceedings of the CSU-EPSA, 2019,43(2): 171-178.

[22] Cui L , Yang J , Wang L , et al. Theory and Application of Weak Signal Detection Based on Stochastic Resonance Mechanism[J]. Security and Communication Networks, 2021,32(2):1-9.

[23] Yu F M , Lee K C , Jwo K W , et al. Low Distortion of Noise Filter Realization with 6.34 V/μs Fast Slew Rate and 120 mVp-p Output Noise Signal[J]. Sensors, 2021, 21(3):1008-1011.

[24] Pantke D , Mueller F , Reinartz S , et al. Frequency-selective signal enhancement by a passive dual coil resonator for magnetic particle imaging. 2022, 241(6):108-113.

[25] Zhou X , Sun Z , Wu H . Wireless Signal Enhancement Based on Generative Adversarial Networks[J]. Ad Hoc Networks, 2020, 103(8):102-110.

[26] Chuang C C , Lee C C , Yeng C H , et al. Convolutional denoising autoencoder based SSVEP signal enhancement to SSVEP-based BCIs[J]. Microsystem Technologies, 2019,32(10):321-325.

[27] Jun-Xia L I , Yuan S F . Signal Enhancement Algorithm of OFDM Network Based on Frequency Domain Narrowband Noise Cancellation[J]. Journal of Southwest China Normal University(Natural Science Edition), 2019, 52(3):83-89.